\documentclass[twocolumn]{aastex631}

\newcommand\add[1]{\textcolor{black}{#1}}
\newcommand\al{$^{26}$Al}

\received{\today}

\submitjournal{ApJ}

\usepackage {threeparttable}
\usepackage{amsmath}
\usepackage{color}

\shorttitle{$^{26}$Al Injection into Protosolar Disk from SN}
\shortauthors{Sawada et al.}

\graphicspath{{./}{figures/}}

\begin{document}

\title{Self-consistent Conditions for $^{26}$Al Injection into Protosolar Disk from a Nearby Supernova}


\author[0000-0003-4876-5996]{Ryo Sawada}
\correspondingauthor{Ryo Sawada}
\email{ryo@g.ecc.u-tokyo.ac.jp}
\affiliation{Department of Earth Science and Astronomy, Graduate School of Arts and Sciences, The University of Tokyo, Tokyo 153-8902, Japan}

\author[0000-0002-6602-7113]{Tetsuo Taki}
\affiliation{Department of Earth Science and Astronomy, Graduate School of Arts and Sciences, The University of Tokyo, Tokyo 153-8902, Japan}

\author[0000-0003-1965-1586]{Hiroyuki Kurokawa}
\affiliation{Department of Earth Science and Astronomy, Graduate School of Arts and Sciences, The University of Tokyo, Tokyo 153-8902, Japan}
\affiliation{Department of Earth and Planetary Science, Graduate School of Science, The University of Tokyo, Tokyo 113-0033, Japan}

\author[0000-0002-7443-2215]{Yudai Suwa}
\affiliation{Department of Earth Science and Astronomy, Graduate School of Arts and Sciences, The University of Tokyo, Tokyo 153-8902, Japan}
\affiliation{Center for Gravitational Physics and Quantum Information, Yukawa Institute for Theoretical Physics, Kyoto University, Kyoto 606-8502, Japan}

\begin{abstract}

The early solar system contained a short-lived radionuclide, $^{26}$Al (its half-life time $t_{1/2} = 0.7$ Myr). The decay energy $^{26}$Al is thought to have controlled the thermal evolution of planetesimals and, possibly, the water contents of planets. Many hypotheses have been proposed for the origin of $^{26}$Al in the solar system. One of the possible hypotheses is the `disk injection scenario'; when the protoplanetary disk of the solar system had already formed, a nearby $(<1 \,\mathrm{pc})$ supernova injected radioactive material directly into the disk. Such a $^{26}$Al injection hypothesis has been tested so far with limited setups for disk structure and supernova distance, and treated disk disruption and $^{26}$Al injection separately. Here, we revisit this problem to investigate whether there are self-consistent conditions under which the surviving disk radius can receive enough $^{26}$Al which can account for the abundance in the early solar system. We also consider a range of disk mass and structure, $^{26}$Al yields from supernova, and a large dust mass fraction $\eta_\mathrm{d}$. We find that $^{26}$Al yields of supernova are required as $\gtrsim 2.1\times10^{-3}M_\odot(\eta_\mathrm{d}/0.2)^{-1}$, challenging to achieve with known possible $^{26}$Al ejection and dust mass fraction ranges. Furthermore, we find that even if the above conditions are met, the supernova flow changes the disk temperature, which may not be consistent with the solar-system record. Our results place a strong constraint on the disk injection scenario. Rather, we suggest that the fresh $^{26}$Al of the early solar system must have been synthesized/injected in other ways.

\end{abstract}



\section{Introduction} \label{sec:intro}

Short-lived radionuclides (SLRs) with half-lives of tens of Myr or less drive the thermal evolution of planetesimals ($\sim$km-sized bodies in planetary systems) and, consequently, influence the formation of planetary systems \citep[e.g.,][]{2022arXiv220311169D}. Among SLRs, \al\, \citep[its half-life time $t_{1/2}\approx 0.72\ \mathrm{Myr}$,][]{2009E&PSL.287..453A} is thought to be the dominant heat source in the early solar system \citep{1955PNAS...41..127U}. Meteorite analysis (measurements of the daughter nuclide, $^{26}$Mg) provided direct evidence and abundance estimate of \al\, at the time of the formation of calcium--aluminum-rich inclusions (CAIs), the oldest material condensed in the early Solar System \citep{1974Natur.251..495G,1976GeoRL...3...41L}. Depending on the timing of accretion and resulting \al\, content of each body, the radioactive decay heat led to various physicochemical evolution of planetesimals, including core-mantle differentiation of achondrite parent bodies \citep[e.g.,][]{2019E&PSL.507..154L} and aqueous chemistry to form secondary minerals \citep[e.g.,][]{2022AGUA....300568K} and to drive chemical evolution of organics \citep[][]{2022NatCo..13.4893L}. The radioactive decay heating of planetesimals and subsequent water loss has been proposed to have controlled water budgets of planets and their diversity in the solar and extrasolar systems \citep{2015ApJ...804....9C,2019NatAs...3..307L}.

Moreover, SLRs provide information on the astrophysical conditions for star formation and the birth environment of the solar system.
The discovery of undecayed \al\,(hereafter referred as `fresh \al') in the protosolar disk at the time of CAIs formation \citep{1976GeoRL...3...41L} led to three origin theories: inheritance, irradiation, and injection. 
The inheritance hypothesis suggests \al\, came from the parent giant molecular cloud, enriched by star formation in spiral arms \citep[e.g.,][]{2018MNRAS.480.4025F}. However, this theory faces a timing issue: molecular clouds forming the solar system take over $1$ Myr, during which \al\, would decay, conflicting with meteorite data. 
The irradiation hypothesis proposes that \al\, is synthesized by irradiation of accelerated particles, such as from solar flares, during solar system formation.  
This model explains $^{10}$Be and also some \al\, in CAIs \citep{2017NatAs...1E..55S,2019A&A...624A.131J}, but it is unclear whether the appropriate mechanism for producing accelerated particles can be explained from astronomical sources.\footnote{
The `X-wind model' of \citet{2001ApJ...548.1029S} has long been a leading example of a method to explain {\it in situ} SLRs synthesis, but it has faced critical scrutiny \citep{2010ApJ...725..692D}.}
Thus, the long-standing and most reliable approach is the injection hypothesis: 
if nearby core-collapse supernova (hereafter we refer as `supernova (SN)') ejecta are injected into the protosolar system, 
it is known that the relative fraction of injected SLRs is roughly consistent with the abundances found in meteorites 
\citep[excluding $^{10}$Be; e.g.,][]{2000SSRv...92..133M}.

In the injection hypothesis, two major injection timing models have been considered. 
One is that the SN ejecta enters the molecular cloud core, leading to its collapse from the SN shock wave \citep[e.g.,][]{1977Icar...30..447C,1998ApJ...494L.103B,2008ApJ...686L.119B}, and the other is that the SN ejecta is injected directly into the already formed protosolar disk \citep[e.g.,][]{2007ApJ...662.1268O,2010ApJ...711..597O,2021ApJ...908...64F}.
\add{Also mentioned is the injection into filaments, which assumes both cases \citep{2023ApJ...947L..29A}.}

For the case of direct injection into the protosolar disk, several previous studies have investigated the pressure conditions under which the disk can be injected without disruption. 
\citet{2000ApJ...538L.151C} analytically estimated the disruption of the protosolar disk by SN shocks and found that, under typical conditions, the disk could be partially stripped but not completely destroyed.
\citet{2005ASPC..341..527O} also estimated that for a disk with mass $M_\mathrm{disk}\approx 0.01 M_\odot$ and radius  $R\approx30\,\mathrm{au}$, efficient capture of SN ejecta from a $d\approx 0.3 \,\mathrm{pc}$ explosion  would explain the estimated \al\, amount from the meteorite.
Hydrodynamical simulations by \citet{2007ApJ...662.1268O} 
\add{(and also \citet{2017MNRAS.469.1117C}, \citet{2018A&A...616A..85P}, and \citet{2019A&A...622A..69P})}
show that the disk resists total destruction by SN impacts, but the contribution of gas-phase ejecta for injection is minimal, less than 1 \%. 
However, \citet{2010ApJ...711..597O} showed that, while small dust grains ($\lesssim 0.1\ \mathrm{\mu m}$) follow the gas and not injected into the disk, large grains ($\gtrsim 1\ \mathrm{\mu m}$) are efficiently injected with $\sim 100\%$ efficiency.
If sufficient \al\, is contained in large dust grains condensed from supernova ejecta, the protosolar disk would receive enough \al\, to account for the meteorite ratio. 
Thus the injection of \al\, into the disk would depend on the ability to inject SN large dust grains.

Here, we raise the question of whether large dust grains can inject a sufficient amount of \al, accepting a certain amount of disk disruption. 
The previous analytical studies have treated the structure in one-zone model \citep{2000ApJ...538L.151C,2005ASPC..341..527O}, and previous numerical studies covered only a limited number of disk models \citep{2007ApJ...662.1268O,2010ApJ...711..597O}, and investigated the pressure conditions without disruption. 
However, to consider this question, a wide range of disk masses and structures should be considered. 
Also, all of the previous studies focused only on the pressure condition and did not consider the effect of SN on the disk temperature.

Furthermore, recent advancements in observational astronomy have greatly enhanced our understanding of SN explosions in the last decade. 
Notably, the recent direct detection of several progenitors suggests that the majority of massive stars above $\sim20M_\odot$ may collapse quietly to black holes and that the explosions remain undetected \citep[e.g.,][]{2015PASA...32...16S, 2017MNRAS.468.4968A}.
Recent numerical calculations have also reported results leading to implosion at progenitor masses above $\gtrsim20M_\odot$ \citep[e.g.,][]{2012ApJ...757...69U,2016ApJ...821...38S}, making the $25M_\odot$ supernova model, at least as frequently cited in previous discussions of early solar system composition \citep{2005ASPC..341..527O,2007ApJ...662.1268O}, inappropriate as a typical value.
In addition, over the past decade, there has been much discussion of the indeterminacy of the \al\, yield based on a more modern understanding of stellar evolution and supernova explosions \citep[e.g.,][]{2007PhR...442..269W,2010ApJ...718..357T,2019ApJ...884...38B,2021ApJ...923...47B,2023ApJ...951..110B}.
Observations of dust mass abundance ratios, also important in this study, are now available for supernovae with younger timescales of interest in this study \citep[$\sim40$ yr; e.g., SN 1987A and SN 1980K,][]{2011Sci...333.1258M,2023arXiv231003448Z}.

\begin{figure*}[htpb]
\centering
  \includegraphics[width=0.95\textwidth]{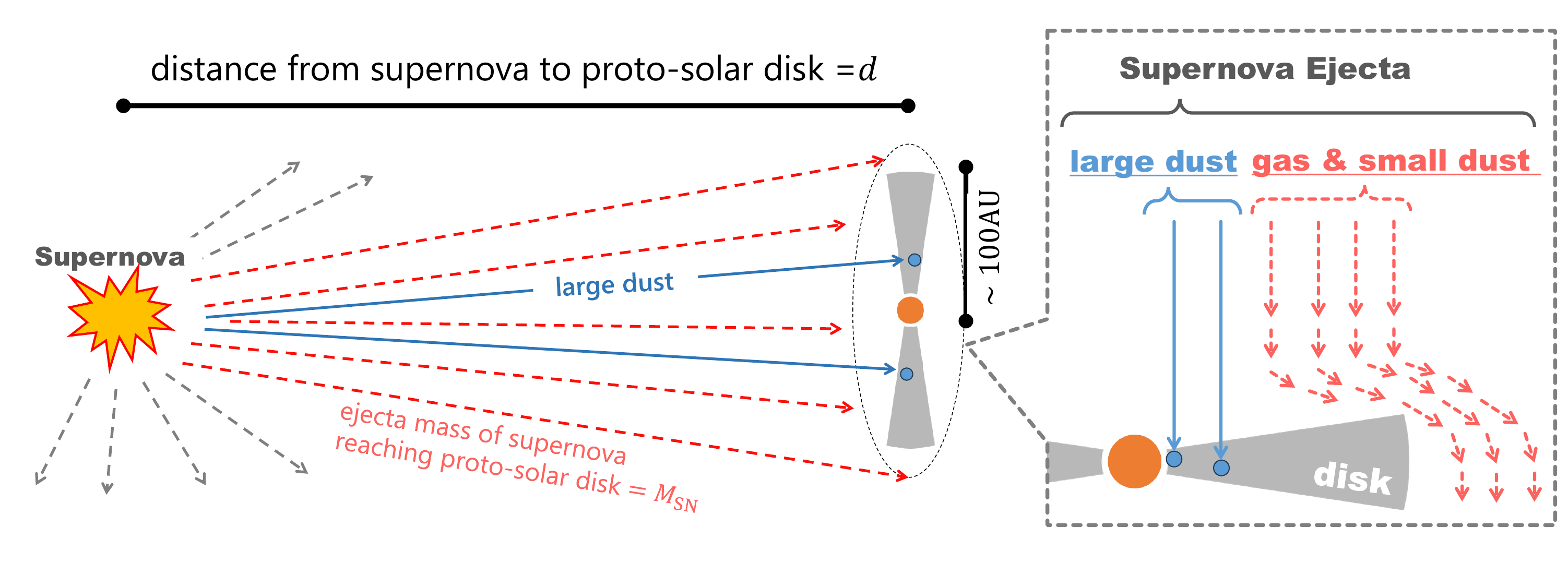}
    \caption{Schematic picture of supernova dust containing \al\, being injected directly into an existing protosolar disk a few tenths of a parsec away.   }
\label{fig:schematic}
\end{figure*}

To summarize, we assume that \al\, from a SN, is injected into an already formed protosolar disk (Figure \ref{fig:schematic}) and investigate the conditions under which the surviving disk radius can capture a sufficient amount of \al\, for planet formation while allowing some disruption of the disk.
We consider a diversity of disk mass and its structure, \al\, yields of SN, and large-dust mass fraction of \al.
We found that for all supernova models with progenitor masses of $11-40 M_\odot$, the disk radius is disrupted to less than 30 au if a sufficient amount of injection is achieved.
Thus we conclude that even if partial disk disruption is allowed, the hypothesis of \al\, injection into an already formed protosolar disk cannot be reproduced in almost all supernova explosion models.

The paper is organized as follows: Section \ref{sec:model} details the system setup and \al\, injection conditions. Section \ref{sec:press} examines conditions for disk disruption by supernova flows. Section \ref{sec:criteria} addresses disk structure and mass, determining conditions for sufficient \al\, capture. Section \ref{sec:thermal} explores supernova impacts on disk temperature and planet formation. Section \ref{sec:discussion} discusses limitations of our study. Finally, Section \ref{sec:summary} summarizes our findings.

\section{Basic picture}  \label{sec:model}

In this section, some definitions are summarized below to clarify our model.  
Figure \ref{fig:schematic} shows a schematic picture of this study, and Table \ref{tb:variables} shows the definitions of the variables in our study.
The three major assumptions in this study are: 
(1) SLRs are injected from a nearby supernova only after the protosolar disk is formed. 
(2) All of the $^{26}$Al in the protosolar disk is supplied by a nearby supernova.
(3) For simplicity, the protosolar disk is in face-on contact with the supernova ejecta.

As background for each assumption, (1) at least one supernova, which would directly inject the SLRs into the early solar system, is expected to occur in the Sun's birth cluster during the duration of CAI formation \citep[e.g.,][]{2023A&A...670A.105A}.
And (2) since the timescale for disk formation is sufficiently long compared to the lifetime of \al\,, one would expect that all of the \al\, originally contained in the disk, has decayed and can be neglected.
\add{Finally, 
(3) the assumption of a face-on disk gives upper limits on both the disk disruption and injection efficiencies. 
In practice, the disk is likely to be inclined at an angle, but the face-on disk is the most injectable assumption 
(see Section \ref{subsec:incl} for more detail).
Moreover, as discussed in \citet{2017A&A...604A..88W}, when the disk contacts the external outflow, the disk can be readjusted to be perpendicular to the flow.
Therefore, the inclination angle will not change the overall conclusion.}

\begin{table*}[]
\begin{center}
\caption{Summary of some important variables index}\label{tb:variables}
    \begin{tabular}{p{2.0cm}|p{1.0cm}|p{11.0cm}|p{2cm}}
        \hline \hline
        Name \,&[unit]& description & typical value\\  \hline
        $M_\mathrm{SN,tot}$\,&$[M_\odot]$ & 
        Total ejecta mass of a nearby supernova explosion & 
        $10$\\ \hline
        $M_\mathrm{SN}$\,&$[M_\odot]$ & 
        Mass of the supernova ejecta in contact with the protosolar disk  (referred to as the `contact mass'), assuming an isotropic explosion from a distance of $d$. &  
        $\left(\sim 10^{-6}\right)$\\ \hline
        $M_\mathrm{SN}(\mathrm{^{26}Al})_\mathrm{tot}$\,&$[M_\odot]$ &
        Ejected masses of \al\, from supernova &
        $1.3\times 10^{-4}$\\ \hline 
        $d$\,&$[\mathrm{pc}]$ & 
        Distance from the protosolar disk to the supernova. & 
        $0.1$\\ \hline
        $\eta_\mathrm{d}$\,&$[-]$ & 
        Mass fraction of large dust $(a > 1 \mathrm{\mu}m)$ in supernova ejecta &
        $0.20$\\ \hline
        $M_\mathrm{disk}$\,&$[M_\odot]$ & 
        Mass of the protosolar disk &
        $\sim0.017$\\ \hline
        $R$\,&$[\mathrm{au}]$ & 
        Radius of the protosolar disk &
        $\sim100$\\ \hline
        $\Sigma_\mathrm{disk}$\,&$[\mathrm{g\,cm^{-2}}]$ & 
        Surface density of the protosolar disk &
        $\sim1.7$\\ \hline
        $q$\,&$[-]$ & 
        Radius dependence of the disk surface density (Eq. \eqref{eq:dens-surf})&
        $1-3/2$\\ \hline
        $t_\mathrm{delay}$ & [Myr] &
        Time delay from injection to CAIs formation. &
        $-$
        \\ \hline \hline
    \end{tabular}
\end{center}
\tablecomments{
    The `typical value' is the value adopted in this study to estimate the value.
    And `$[-]$' denotes dimensionless variables. 
    Those marked with `()' depend on other variables, see text.} 
\end{table*}

We first discuss the amount of $\mathrm{^{26}Al}$ injected into the protosolar disk from nearby SN.
To avoid confusion, the mass of the SN ejecta in contact with the protosolar disk is called the `contact mass'  $M_\mathrm{SN}$. 
The relation of this value to the total mass of the SN ejecta $M_\mathrm{SN,tot}$ is as follows
\begin{align}
    &M_\mathrm{SN} 
    \sim \left(\cfrac{\pi R^2}{4\pi d^2}\right)M_\mathrm{SN,tot} \\
    &\approx  6\times 10^{-5}\, M_\odot
    \left(\cfrac{M_\mathrm{SN,tot}}{10\,M_\odot}\right)
    \left(\cfrac{d}{0.1\,\mathrm{pc}}\right)^{-2}
    \left(\cfrac{R}{100\,\mathrm{au}}\right)^2~,\nonumber
\end{align}
where $d$ is the distance from the protosolar disk to the supernova, and $R$ is the radius of the protosolar disk.
The contact mass is composed of gas and dust, and we define the mass fraction of large dust (here defined as grains sufficiently large to be injected into the disk; typically $>1\, \mathrm{\mu m}$ \citealt{2010ApJ...711..597O}) as $\eta_\mathrm{d}$.
Assuming that the $^{26}$Al injection via gas and small dust grains are negligible, we can write the injected \al\, mass as,
\begin{align}
    &M_\mathrm{inje}(\mathrm{^{26}Al})
    \sim \eta_\mathrm{d} \cdot \left(\cfrac{\pi R^2}{4\pi d^2}\right) \cdot M_\mathrm{SN}(\mathrm{^{26}Al})_\mathrm{tot} \nonumber \\ 
    &\approx  1.5\times10^{-10} M_\odot 
    \left(\cfrac{d}{0.1\,\mathrm{pc}}\right)^{-2}
    \left[
    \left( \cfrac{\eta_\mathrm{d}}{0.20} \right)
    \left( \cfrac{ M_\mathrm{SN}(\mathrm{^{26}Al})_\mathrm{tot} }{1.3\times 10^{-4}M_\odot} \right)
    \right] ~,
    \label{eq:al-sn}
\end{align}
where $M_\mathrm{SN}(\mathrm{^{26}Al})_\mathrm{tot}$ is the ejected mass of \al\, from the supernova, with a adopted value of about $\sim 1.3\times 10^{-4} M_\odot$ per event \citep[e.g.,][and see Table \ref{tb:26al}]{1995ApJ...449..204T,2006ApJ...647..483L,2016ApJ...821...38S,2018ApJS..237...13L}. 
There are still many uncertainties regarding the nucleosynthesis of \al\, in supernova, and we listed the theoretical \al\, yields of several different supernova models in Table  \ref{tb:26al}.
Since the typical progenitor mass of recently observed supernovae is $M\approx 8-17M_\odot$ \citep{2015PASA...32...16S}, 
the typical value of $\sim 1.3\times 10^{-4} M_\odot$ used in this study is considered a sufficiently robust upper limit.
Here, we adopted the large dust mass fraction $\eta_\mathrm{d} \sim 20\%$ as the typical value \citep[e.g., SN 1987A;][and also see Section \ref{subsec:large-dust}]{2011Sci...333.1258M}.
Note that the timescale from the ejection of the supernova to its injection into the disk is $t_\mathrm{SN}\sim d/v_\mathrm{SN}\sim 43\,\mathrm{yr}(d/0.1\,\mathrm{pc})(E_\mathrm{exp}/10^{51}\,\mathrm{erg})^{-1/2}$, so that the radioactive decay of \al\, during the travel is negligible.

\begin{table*}[]
\begin{center}
\caption{Summary of theoretical \al\, yield of 5 different supernova models. All values in the table are in $M_\odot$}\label{tb:26al}
    \begin{tabular}{c|c|c|cc|c}
    Progenitor Mass & WW95$^{(1)}$ 
    & LC06$^{(2)}$
    & LC18$^{(3-1)}$
    & LC18$^{(3-2)}$
    & S16$^{(4)}$
    \\ \hline
    \multicolumn{6}{c}{\footnotesize Standard-mass progenitor models supported by recent observations.} \\ \hline
    11 & $1.7\times10^{-5}$ & $1.6\times10^{-5}$ & - & - & $1.1\times10^{-5}$\\
    12 & $2.0\times10^{-5}$ & $2.1\times10^{-5}$ & - & - & $1.2\times10^{-5}$\\
    13 & $2.8\times10^{-5}$ & $2.4\times10^{-5}$ & $1.9\times10^{-5}$ & $4.4\times10^{-5}$ & $1.7\times10^{-5}$\\
    15 & $4.3\times10^{-5}$ & $1.3\times10^{-4}$ & $3.8\times10^{-5}$ & $6.9\times10^{-5}$ & $3.2\times10^{-5}$\\ 
    \hline
    \multicolumn{6}{c}{\footnotesize  High-mass progenitor models with many observational uncertainties.} \\ 
    \hline
    20 & $3.5\times10^{-5}$ & $5.4\times10^{-5}$ & $6.3\times10^{-5}$ & $5.4\times10^{-5}$ & ${}^\dagger\,1.6\times10^{-7}$\\ 
    25 & $1.3\times10^{-4}$ & $8.6\times10^{-5}$ & $7.6\times10^{-5}$ & $1.3\times10^{-4}$ & ${}^\dagger\,1.9\times10^{-6}$\\
    30 & $2.7\times10^{-4}$ & $9.9\times10^{-5}$ & $3.7\times10^{-6}$ & $2.1\times10^{-5}$ & ${}^\dagger\,8.6\times10^{-6}$\\
    35 & $3.5\times10^{-4}$ & $8.4\times10^{-5}$ & - & - & ${}^\dagger\,2.3\times10^{-5}$\\
    40 & $3.6\times10^{-4}$ & $1.2\times10^{-4}$ & $1.2\times10^{-5}$ & $3.9\times10^{-5}$ & ${}^\dagger\,3.4\times10^{-5}$\\
    \hline \hline
    \end{tabular}
\end{center}
\tablecomments{(1) \citet{1995ApJS..101..181W}, (2)\citet{2006ApJ...647..483L}, (3-1,2) \citet{2018ApJS..237...13L}. (4) \citet{2016ApJ...821...38S}.
All models assume a non-rotating progenitor model with solar metallicity, except for the rotating star model $(v=300\,\mathrm{km\,s^{-1}})$ in (3-2).
The supernova model in (4) considers the possibility of explosion, and the nucleosynthesis calculation is self-consistent with it, so there are cases of explosion failure, in which case the tagger symbol ‘$(\dagger)$’ is added in the table.
The other models in (1)-(3) are based on artificial explosions, regardless of the explodability.
} 
\end{table*}

Next, from meteorite measurements, we know short-lived nuclide/stable isotope abundance ratios at the time of CAI formation as $N_\mathrm{r}/N_\mathrm{s}=^{26}\mathrm{Al}/^{27}\mathrm{Al}\approx5.23\times10^{-5}$ \citep[e.g.,][]{2008E&PSL.272..353J}, which are extrapolated to the time of the formation of CAIs. 
Using this ratio and the mass fraction of $^{27}$Al in the solar abundance $X_\odot(\mathrm{^{27}Al})\approx5\times10^{-5}$ \citep{2009ARA&A..47..481A}, the total amount of undecayed \al\, in the protosolar disk $M_\mathrm{disk}(\mathrm{^{26}Al})$ can also be written as
\begin{align}
    &M_\mathrm{disk}(\mathrm{^{26}Al})
    \sim \cfrac{ \left( ^{26}\mathrm{Al}/^{27}\mathrm{Al} \right)\cdot
    X_\odot(\mathrm{^{27}Al})\cdot M_\mathrm{disk} }
    {\exp{(-t_\mathrm{delay}/1.0\,\mathrm{Myr})} }\nonumber \\
    &\gtrsim  4.3\times 10^{-11} M_\odot
    \left( \cfrac{M_\mathrm{disk}}{0.017M_\odot} \right)  
    ~,
    \label{eq:al-disk}
\end{align}
where the exponential function in the equation is the correction factor between the actual injected mass and the observed mass due to the time delay $t_\mathrm{delay}$ from injection to CAIs formation.
In this study, we have chosen a value of $t_\mathrm{delay}=0$. Although this term is not insignificant, it represents the minimum amount that should be introduced to give robust conditions.
Here, to satisfy $ M_\mathrm{disk}(\mathrm{^{26}Al})= M_\mathrm{inje}(\mathrm{^{26}Al})$,
we obtain the following conditions for the distance of the supernova:
\begin{align}
    d \lesssim 0.20\,\mathrm{pc} 
    &\left[
    \left( \cfrac{\eta_\mathrm{d}}{0.20} \right)
    \left( \cfrac{ M_\mathrm{SN}(\mathrm{^{26}Al})_\mathrm{tot} }{1.3\times 10^{-4}M_\odot} \right)
    \right]^{1/2} \nonumber \\
    &\times\left( \cfrac{M_\mathrm{disk}}{0.017M_\odot} \right)^{-1/2} 
    \left(\cfrac{R}{100\,\mathrm{au}}\right) ~.
    \label{eq:d-26al}
\end{align}
where the disk radius $R$ only determines the solid angle at which \al\, can be received from SN, once $100$ au is adopted. Similarly, the disk mass $M_\mathrm{disk}$ here normalizes the total amount of \al\, required, and we once adopted $0.017M_\odot$.

The injectable distance obtained from Eq. \eqref{eq:d-26al} is the same as the distance obtained by matching the parameter $\eta_\mathrm{d}$ with previous studies:
\cite{2005ASPC..341..527O} assumed a disk radius of $R=30\,\mathrm{au}$, the injection rate of 100\%, and chose a supernova that can eject more \al\, than that in a typical supernova ($40M_\odot$ progenitor model for \citealt{1995ApJS..101..181W}, assuming $\sim 3.6\times10^{-4}M_\odot$ yields) and derived the distance $d\approx 0.3\,\mathrm{pc}$.
Also, \cite{2007ApJ...662.1268O} suggested that \al\, is insufficient at $d=0.1\,\mathrm{pc}$ for the 4\% injection rate suggested from hydrodynamical simulation.
These estimates are consistent with our result obtained from Eq. \eqref{eq:d-26al}, while we note that gas and dust are not distinguished in these estimates, and thus the correspondence to $\eta_\mathrm{d}$ values in our model has no physical meaning on the actual $\eta_\mathrm{d}$ value.

\section{Pressure stability condition}  \label{sec:press}

\subsection{Basic picture of disk disruption}
\begin{figure}
\centering
  \includegraphics[width=0.45\textwidth]{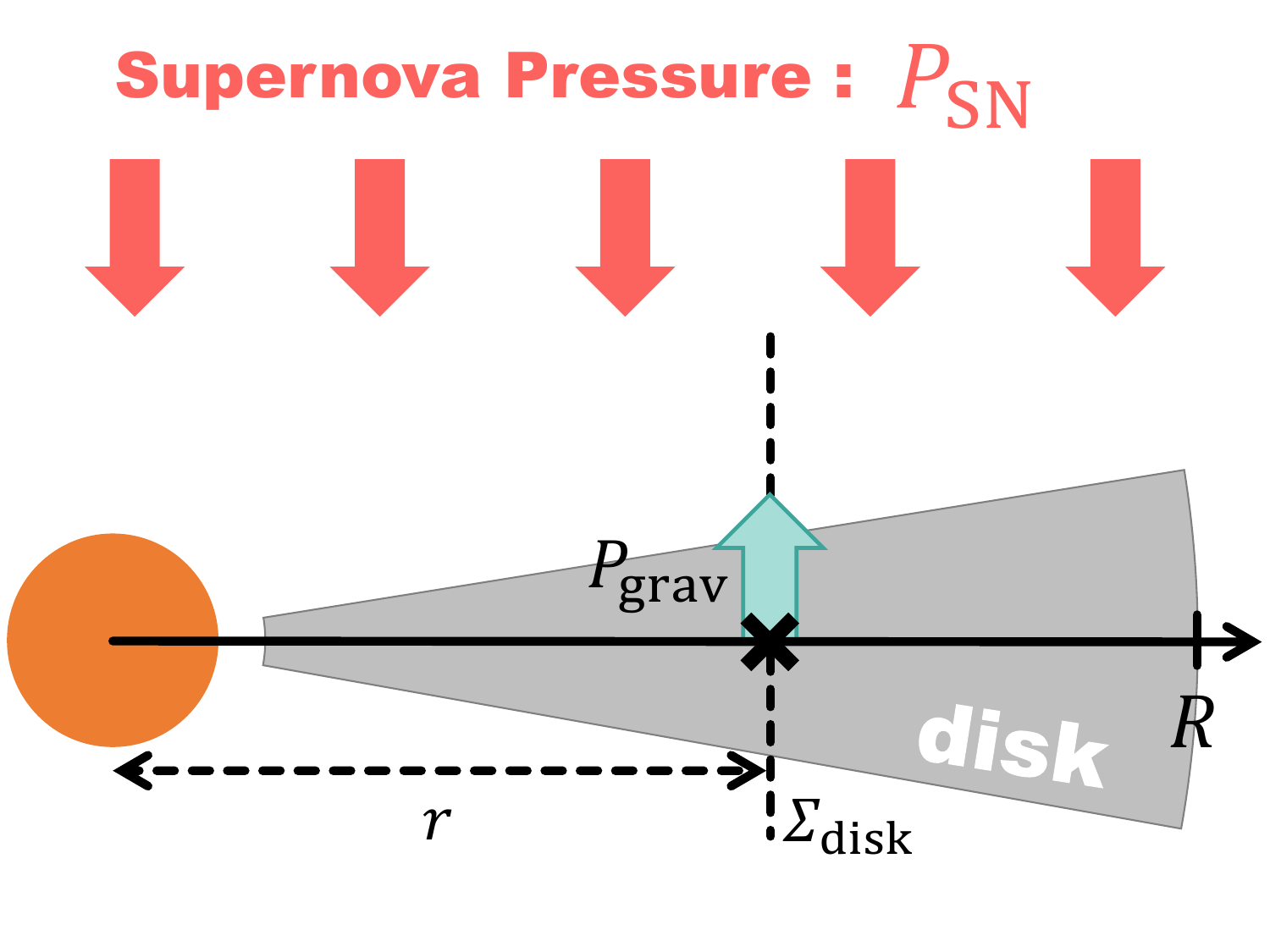}
    \caption{
    Schematic picture considering whether the disk disruption is caused by ram pressure from a supernova flow.
    }
\label{fig:press}
\end{figure}

Next, we discuss the stability of the protosolar disk due to dynamical pressure as shown in Figure \ref{fig:press}. 
Since it is straightforward to speculate wether the ram pressure of the dense SN ejecta $(n\sim10^3\,\mathrm{cm}^{-3}\left(d/1\,\mathrm{pc}\right)^{-3})$ can destroy the protosolar disk, it has been investigated analytically/numerically in many previous studies \citep[e.g.,][]{2000ApJ...538L.151C,2007ApJ...662.1268O,2017MNRAS.469.1117C}.
Our conclusion that the disk can survive the supernova at distances $\gtrsim\mathcal{O}(0.1)\,\mathrm{pc}$ is the same as in previous studies. 
The pressure stability condition derived below is almost identical to that in \citet{2000ApJ...538L.151C}, but here we briefly review the condition for completeness of the discussion.

We consider that the disk is disrupted if the ram pressure of the supernova flows $P_\mathrm{sn}$ exceeds the gravitational force per unit area, $P_\mathrm{grav}$, which keeps the disk bound to the central pre-main-sequence star. An estimate of the gravitational force per unit area is
\begin{align}
    P_\mathrm{grav} 
    &= \cfrac{GM_\star\Sigma_\mathrm{disk}}{r^2}\nonumber\\
    &\approx 1\times10^{-4} \,\mathrm{dyn\,cm^{-2}} 
     \left(\cfrac{\Sigma_\mathrm{disk}}{1.7\,\mathrm{g\,cm^{-2}}}\right)    
     \left(\cfrac{r}{100\,\mathrm{au}}\right)^{-2}~,    
\end{align}
where $M_\star$ is the mass of the central star and adopted as $M_\star=1M_\odot$, $r$ is the distance of the radial direction from the center, and $\Sigma_\mathrm{disk}$ is the surface
density at $r$. 
Here, typical value of the disk surface density $\Sigma_\mathrm{disk}$ and radius $r$ were adopted once from the minimum mass solar nebula model. 
For a more detailed discussion, the values will be redefined in Section \ref{subsec:disk}.

Next, we calculate the ram pressure that the disk receives from the freely expanding supernova ejecta.
For simplicity, we use the thin-shell approximation for the ejecta \citep[e.g.,][]{1969JFM....35...53L,1990ApJ...354..513K}.
In this scenario, we consider an isotropic ejecta, which forms a hot bubble, and the pressure of this bubble is given by, 
\begin{align}
    P_\mathrm{SN} 
    &\approx \rho_\mathrm{SN}v_\mathrm{SN}^2
    =\cfrac{2E_\mathrm{expl}}{4\pi d^2 \Delta d}~ \nonumber\\
    &\approx 7\times 10^{-2}\,\mathrm{dyn\,cm^{-2}} 
    \left(\cfrac{E_\mathrm{expl}}{10^{51}\,\mathrm{erg}}\right)
    \left(\cfrac{d}{0.1\,\mathrm{pc}}\right)^{-3}~,
\end{align}
where $E_\mathrm{expl}$ is the explosion energy of SN. $\Delta d$ is the thickness of the shell, which is $\Delta d/d=1/12$ in the case of the thin shell for an ideal gas \citep{1969JFM....35...53L}.

Here, the stability criterion $P_\mathrm{grav}> P_\mathrm{SN}$ for the disk region outside the radius $r$ is given by,
\begin{align}
    d > & 0.86\,\mathrm{pc}
    \left(\cfrac{E_\mathrm{expl}}{10^{51}\,\mathrm{erg}}\right)^{1/3}
    \left(\cfrac{\Sigma_\mathrm{disk}}{1.7\,\mathrm{g\,cm^{-2}}}\right)^{-1/3}    
    \left(\cfrac{r}{100\,\mathrm{au}}\right)^{2/3}
    ~. \label{eq:d-press}
\end{align}
The disk disruption distance obtained from Eq. \eqref{eq:d-press} is the same as the distance obtained from the disk surface density and radius in previous studies:
\cite{2000ApJ...538L.151C} assumed a disk radius of $R=10^{15}\,\mathrm{cm}\approx66\,\mathrm{au}$ and a uniform density distribution $\Sigma_\mathrm{disk}=M_\mathrm{disk}/(\pi R^2)\approx6.4\,\mathrm{g\,cm^{-2}}$, and also used a uniform density for supernova $(\Delta d\to d)$, so its gravitational force is $P_\mathrm{grav}\approx10^{-3}\,\mathrm{dyn\,cm^{-2}}$ and the stable criteria becomes $d\gtrsim 0.25\,\mathrm{pc}$.

\subsection{Formulation with disk structure assumption} \label{subsec:disk}

Here, based on the minimum mass solar nebula model \citep[e.g.,][]{1981PThPS..70...35H,2010apf..book.....A}, we adopt the following disk model for a broad range of disk masses and structures:
\begin{align}
    \Sigma_\mathrm{disk}=10\,\mathrm{g\,cm^{-2}}\,\left(\cfrac{r}{\mathrm{30\,au}}\right)^{-q}
    \left(\cfrac{M_\mathrm{disk}}{0.017\,M_\odot}\right)
    ~,\label{eq:dens-surf}
\end{align}
where $q$ is the radius dependence of the disk surface density, $1< q \leq  3/2$.
For instance, $q =3/2$ in the minimum-mass solar-nebula model \citep{1981PThPS..70...35H} while $q=1$ in a viscous accretion disk model with a constant turbulent alpha parameter \citep{2010apf..book.....A}.
In this case, the radius $R$ of the proto-solar disk in Section \ref{sec:model} is then defined as
\begin{align}
   M_\mathrm{disk}
   & =\int_0^R 2\pi r \Sigma_\mathrm{disk} dr 
   ~,\label{eq:dens-surf-R}
\end{align}
and we can rewrite the disk radius $R$ as
\begin{align}
    R \approx 30\,\mathrm{au} \,
    \times(2.7\times(2-q))^{\frac{1}{2-q}}
   ~.\label{eq:dens-surf-R}
\end{align}
For $q=1$, $R\approx 81\,\mathrm{au}$ is obtained from Eq.\eqref{eq:dens-surf-R}. Also, when $q=3/2$, $R\approx 55\,\mathrm{au}$ is obtained.

Then, the stability criterion of Eq.\eqref{eq:d-press} can be rewritten as
\begin{align}
    d > & 0.21\,\mathrm{pc}
    \left(\cfrac{E_\mathrm{expl}}{10^{51}\,\mathrm{erg}}\right)^{1/3}
    \left(\cfrac{r}{30\,\mathrm{au}}\right)^{\frac{2+q}{3}}
    \left(\cfrac{M_\mathrm{disk}}{0.017\,M_\odot}\right)^{-1/3}
    ~. \label{eq:d-press-q}
\end{align}


\section{Conditions where consistent distances exist}
\label{sec:criteria}
Here, a simple comparison of the \al\, injection distance (Eq. \eqref{eq:d-26al}) and the disk disruption distance (Eq. \eqref{eq:d-press-q}) shows that the answer is not clear and is a very complicated problem to discuss.
Therefore, we here treat this as a more realistic problem by investigating whether a sufficient \al\, injection can be achieved while allowing some disk disruption.
In this section, we describe how to solve this.

To solve this problem, we first define the two radii with assuming the distance from the supernova to the disk:
By rewriting Eq. \eqref{eq:d-press-q} with the disk radius as a function of distance, we obtain the outermost radius $r_\mathrm{surv}$ of the disk that remains unbroken at a given distance $d$ from the SN (we refer to this as the `surviving radius'): 
\begin{align}
    r_\mathrm{surv} 
    =  30\,\mathrm{au}  
    \left(\cfrac{E_\mathrm{expl}}{10^{51}\,\mathrm{erg}}\right)^{\frac{-1}{2+q}}
    &\left(\cfrac{d}{ 0.21\,\mathrm{pc}}\right)^{\frac{3}{2+q}} \nonumber \\
    &\times\left(\cfrac{M_\mathrm{disk}}{0.017\,M_\odot}\right)^{\frac{1}{2+q}}
    \label{eq:surv-radi}~.
\end{align}
Also, by rewriting Eq. \eqref{eq:d-26al} with the disk radius as a function of distance, we also obtain the radius $r_\mathrm{req}$ required to receive a sufficient amount of \al\, for a given distance $d$ from SN (call this the `required radius for \al'):
\begin{align}
    r_\mathrm{req} \approx  120\,\mathrm{au}  
    &\left(\cfrac{d}{ 0.21\,\mathrm{pc}}\right) 
    \left(\cfrac{M_\mathrm{disk}}{0.017\,M_\odot}\right)^{1/2} \nonumber \\
    &\times\left[\left( \cfrac{\eta_\mathrm{d}}{0.20} \right)
    \left( \cfrac{ M_\mathrm{SN}(\mathrm{^{26}Al})_\mathrm{tot} }{1.3\times 10^{-4}M_\odot} \right) 
    \right]^{-1/2}
    \label{eq:req-radi}~.
\end{align}

Then, there are three conditions that must hold for the different radii given by Eqs.\eqref{eq:dens-surf-R}, \eqref{eq:surv-radi}, and \eqref{eq:req-radi} as follows,
\begin{align}
    &\mathrm{(i)}   \qquad 
    r_\mathrm{req} < R~,  \nonumber\\
    &\mathrm{(ii)}  \qquad 
    R_\mathrm{min} < r_\mathrm{surv}~\nonumber,\\
    &\mathrm{(iii)} \qquad 
    r_\mathrm{req} < r_\mathrm{surv}~\nonumber.
\end{align}
The meaning are following: 
(i) 
the disk radius that gives the necessary solid angle for sufficient \al\, injection must be smaller than the radius $R$ estimated from the disk mass in Eq.\eqref{eq:dens-surf-R},
(ii) 
the disk radius that survives the disruption by the supernova flow should remain at least to the Neptune formation radius $R_\mathrm{min} = 30\,\mathrm{au}$,
and (iii) the surviving disk radius $r_\mathrm{surv}$ is more than the required disk radius $r_\mathrm{req}$ that can intercept a sufficient amount of \al.

\begin{figure*}[htbp]
\centering
  \includegraphics[width=0.95\textwidth]{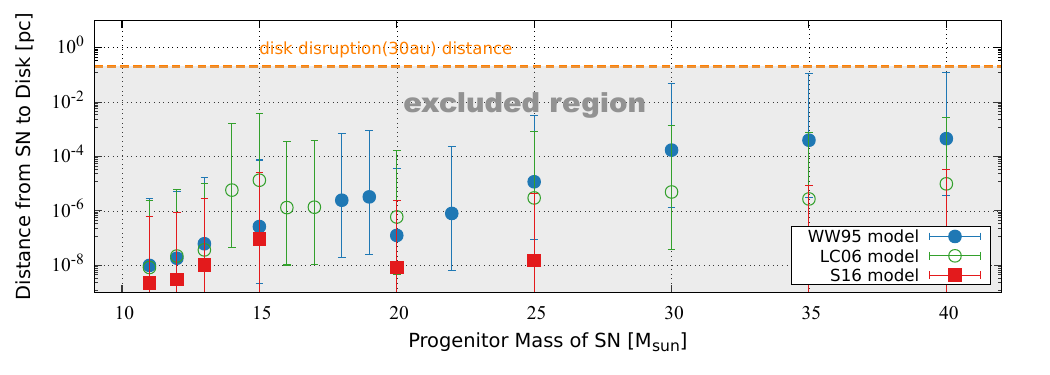}
    \caption{Summary of results for the distance required for \al\, injection for each supernova model and disk disruption condition.
    Results are shown assuming disk mass $M_\mathrm{disk}=0.017M_\odot$ and surface density structure $\Sigma_\mathrm{disk}\propto r^{-3/2}$.
    The orange dotted line indicates the nearest distance where the disk radius remains undisrupted up to $r=30$ au.
    The points indicate the distance at which a sufficient amount of \al\, can be injected into the surviving disk while allowing partial disk disruption in the case that the \al\, yield for each supernova model is adopted (with the large dust mass fraction $\eta_\mathrm{d} =20\%$). Error bars correspond to cases where the large dust mass fraction is from 5\% to 100\%.
    For all supernova models with progenitor masses of $11-40 M_\odot$, 
    we show that the disk radius can be disrupted to less than $30$ au if a sufficient amount of injection is achieved.}
\label{fig:summary}
\end{figure*}
\subsection{Achievement Conditions Assuming Supernova Model}

First, given the \al\, yield in the current supernova model, we investigate whether the disk injection scenario can be achieved.
In this case, solving the condition (iii) (i.e., $r_\mathrm{req} < r_\mathrm{surv}$) gives the achievable distance from the supernova to the disk.
For $q\neq1$, the condition (iii) is satisfied in the following:
\begin{align}
    d <  0.21\,\cdot 4^{\frac{2+q}{1-q}} \cdot\,\mathrm{pc}
    &\left(\cfrac{M_\mathrm{disk}}{0.017\,M_\odot}\right)^{\frac{q}{2-2q}}
    \left(\cfrac{E_\mathrm{expl}}{10^{51}\,\mathrm{erg}}\right)^{\frac{1}{1-q}}
    \nonumber \\
    &
    \times\left[
    \left( \cfrac{\eta_\mathrm{d}}{0.20} \right)
    \left( \cfrac{ M_\mathrm{SN}(\mathrm{^{26}Al})_\mathrm{tot} }{1.3\times 10^{-4}M_\odot}\right) 
    \right]^{\frac{2+q}{2q-2}}
    ~.    \label{eq:surv-dist1}
\end{align}
Figure \ref{fig:summary} shows the achievable distance conditions by substituting the \al\, yield of Table \ref{tb:26al} into Eq. \eqref{eq:surv-dist1}.
Also shown in the figure is the distance conditions for the disk radius to remain above $30$ au, obtained from Eq. \eqref{eq:d-press-q}.
It is shown that for all current supernova models, their injectable conditions lead to mostly destruction with only less than 30 au disk remaining.
From this, we conclude that even if partial disk disruption is allowed, almost all supernova explosion models cannot reproduce the hypothesis of direct \al\, injection into an already formed protosolar disk.

\subsection{More General Achievement Conditions}

\begin{figure}
\centering
  \includegraphics[width=0.45\textwidth]{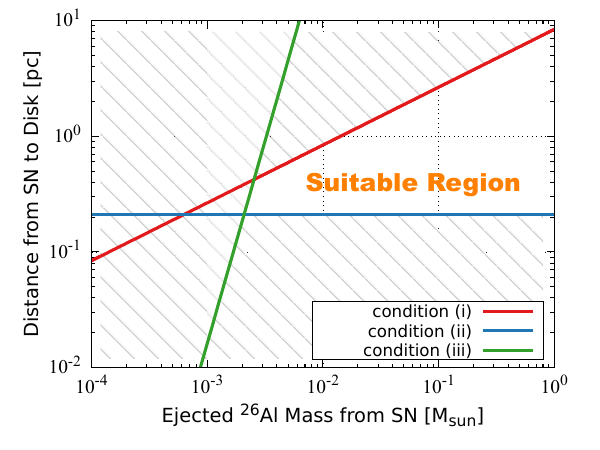}
    \caption{Conditions (i) through (iii) are illustrated on the plane of distance $d$ and \al\, yield.
    Each solid line corresponds to a different condition, in the case of the disk mass $M_\mathrm{disk}=0.017M_\odot$, surface density structure $\Sigma_\mathrm{disk}\propto r^{-3/2}$ and large-dust mass fraction $\eta_\mathrm{d}=20\%$.
    }
\label{fig:region}
\end{figure}

Now, more generally, let us seek what conditions are required for sufficient \al\, injection to be achieved while allowing partial disk disruption.
We can obtain the minimum \al\, yield required to satisfy all the conditions when the equations relating the distance and \al\, yield obtained from each of the conditions (i) through (iii) are coupled.
Figure \ref{fig:region} shows the solution conditions obtained from the coupled equations.

The first two conditions (i) and (ii) are satisfied as follows:
\begin{align}
    \mathrm{(i)}   \,\left(\cfrac{d}{ 0.21\,\mathrm{pc}}\right)< &\,  
    0.25\times(2.7\times(2-q))^{\frac{1}{2-q}}
    \left(\cfrac{M_\mathrm{disk}}{0.017\,M_\odot}\right)^{-1/2} \nonumber \\
    &\times\left[ \left( \cfrac{\eta_\mathrm{d}}{0.20} \right) 
    \left( \cfrac{ M_\mathrm{SN}(\mathrm{^{26}Al})_\mathrm{tot} }{1.3\times 10^{-4}M_\odot} \right)
    \right]^{1/2}
    \\
    \mathrm{(ii)}   \,\left(\cfrac{d}{ 0.21\,\mathrm{pc}}\right)> &\,  
    \left(\cfrac{E_\mathrm{expl}}{10^{51}\,\mathrm{erg}}\right)^{1/3}
    \left(\cfrac{M_\mathrm{disk}}{0.017\,M_\odot}\right)^{-1/3}
    \left( \cfrac{ R_\mathrm{min} }{30\,\mathrm{au} } \right)^{\frac{2+q}{3}}
    ~.
\end{align}
We can confirm that there exists a suitable distance $d$, independent of $q$ (in the range $1\leqq q\leqq3/2$), such that the above two conditions are satisfied.
For $q\neq1$, the condition (iii) is satisfied in the following:
\begin{align}
    \mathrm{(iii)}  \,\left(\cfrac{d}{ 0.21\,\mathrm{pc}}\right)< \,
    &4^{\frac{2+q}{1-q}} 
    \left(\cfrac{M_\mathrm{disk}}{0.017\,M_\odot}\right)^{\frac{q}{2-2q}}
    \left(\cfrac{E_\mathrm{expl}}{10^{51}\,\mathrm{erg}}\right)^{\frac{1}{1-q}}
    \nonumber \\
    &
    \times\left[
    \left( \cfrac{\eta_\mathrm{d}}{0.20} \right)
    \left( \cfrac{ M_\mathrm{SN}(\mathrm{^{26}Al})_\mathrm{tot} }{1.3\times 10^{-4}M_\odot}\right) 
    \right]^{\frac{2+q}{2q-2}}
    ~.    \label{eq:surv-dist}
\end{align}

For there to exist a distance $d$ from the supernova such that all three conditions hold, it is necessary that
\begin{align}
    \left( \cfrac{\eta_\mathrm{d}}{0.20} \right)
    &\left( \cfrac{ M_\mathrm{SN}(\mathrm{^{26}Al})_\mathrm{tot} }{1.3\times 10^{-4}M_\odot} \right)
    \left(\cfrac{E_\mathrm{expl}}{10^{51}\,\mathrm{erg}}\right)^{-2/3}
    \nonumber \\
    &>16
    \left(\cfrac{M_\mathrm{disk}}{0.017\,M_\odot}\right)^{1/3}
    \left( \cfrac{ R_\mathrm{min} }{30\,\mathrm{au} } \right)^{\frac{2q-2}{3}}
    ~.
    \label{eq:result}
\end{align}
Because the disk needs to be sufficiently massive to form the solar system, it is practically difficult to consider a smaller value for the disk mass here.
We can read Eq. \eqref{eq:result} as a requirement for the mass of \al\, contained in the large dust ejected from SN.

The solution is different for the $q=1$ case but agrees with the extrapolated value. 
We attach below a solution for $q=1$ as an example.
By rewriting Eq. \eqref{eq:d-press-q} with the disk radius as a function of distance, we obtain the surviving radius  $r_\mathrm{surv}$ as 
\begin{align}
    r_\mathrm{surv} 
    \approx  30\,\mathrm{au}  
    \left(\cfrac{d}{ 0.21\,\mathrm{pc}}\right)
    \left(\cfrac{M_\mathrm{disk}}{0.017\,M_\odot}\right)^{1/3}
    ~.
\end{align}
The required radius $r_\mathrm{req}$ for \al\, is the same with Eq.\eqref{eq:req-radi}.
In the $q=1$ case, only the condition (iii) to the supernova is given as
\begin{align}
    \left( \cfrac{\eta_\mathrm{d}}{0.20} \right)
    \left( \cfrac{ M_\mathrm{SN}(\mathrm{^{26}Al})_\mathrm{tot} }{1.6\times 10^{-4}M_\odot} \right)
    &
    \left(\cfrac{E_\mathrm{expl}}{10^{51}\,\mathrm{erg}}\right)^{-2/3}
    \nonumber \\
    &>16
    \left(\cfrac{M_\mathrm{disk}}{0.017\,M_\odot}\right)^{1/3}
    ~.
\end{align}
The result is a continuous solution with $q\neq1$.

In summary, the condition (i), (ii), and (iii) are satisfied and the disk injection scenario is realized as 
\begin{align}
    M_\mathrm{SN}(\mathrm{^{26}Al})_\mathrm{tot} 
    \gtrsim 
    2.1\times 10^{-3}M_\odot  
    \left( \cfrac{\eta_\mathrm{d}}{0.20} \right)^{-1} 
    \left(\cfrac{E_\mathrm{expl}}{10^{51}\,\mathrm{erg}}\right)^{2/3}
    ~.\label{eq:result-al}
\end{align}
As can be seen from Table \ref{tb:26al}, there is no supernova model that can achieve such a value, even with a large estimate of the dust fraction $\eta_\mathrm{d}$.
We can also see that this value is not achievable even within the yield diversity that comes from stellar evolutionary processes \citep[see Figure 5 in \citealt{2023ApJ...951..110B} and][]{2019ApJ...884...38B,2021ApJ...923...47B}.
The uncertainty of the nuclear reaction rate also has the potential to increase the \al\, yield, but it seems to be difficult to do so as far as it is currently suggested by \citet{2007PhR...442..269W} and \citet{2010ApJ...718..357T}.

\section{Temperature conditions for disk}  \label{sec:thermal}


\begin{figure}
\centering
  \includegraphics[width=0.45\textwidth]{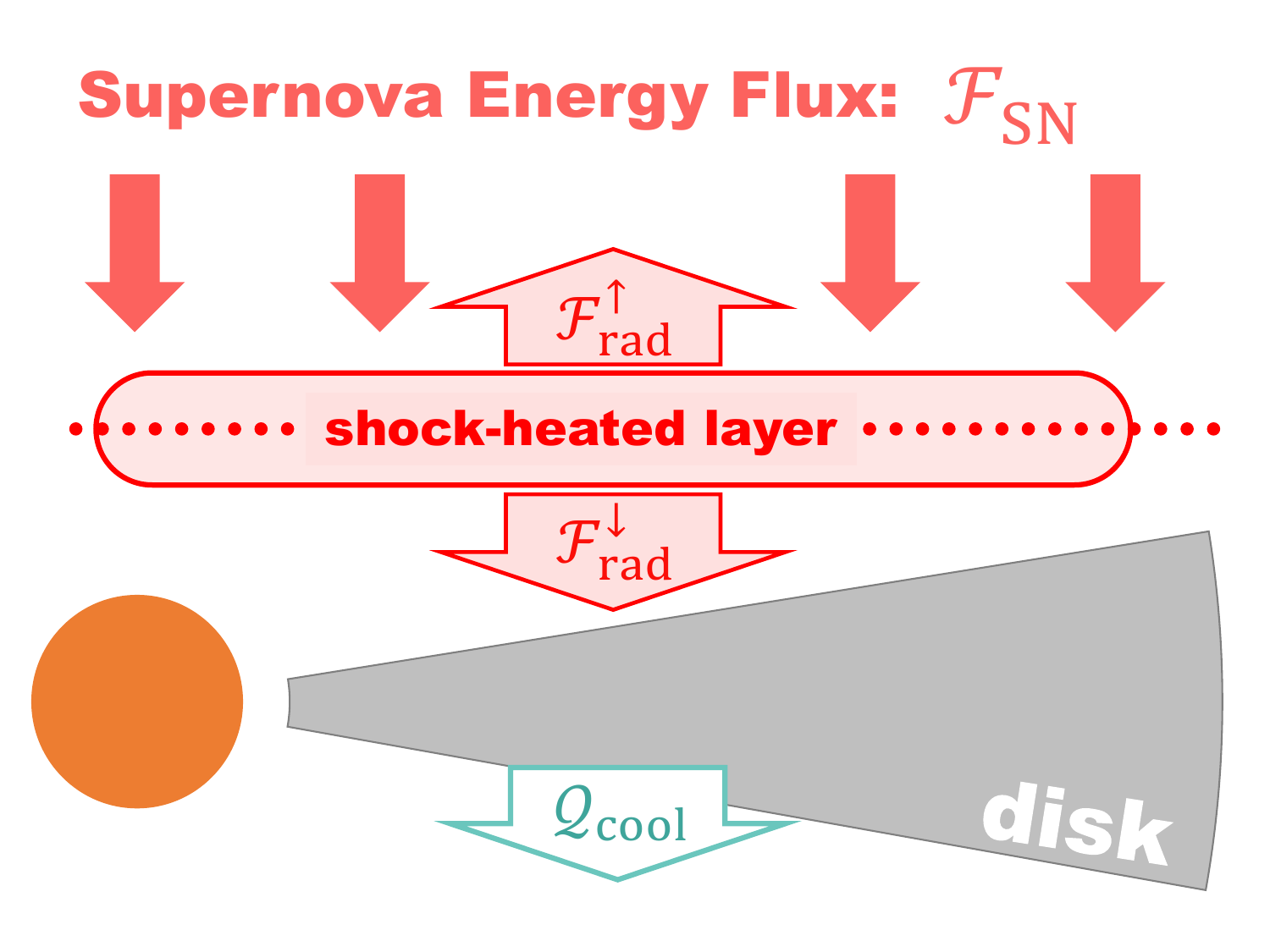}
    \caption{
    Schematic picture of a supernova flow being thermalized at the top of the disk, resulting in heating of the disk.
    }
\label{fig:thermal}
\end{figure}
In this section, we discuss the impact on planet formation if a supernova explosion satisfies the conditions we found in Section \ref{sec:criteria}. 

We consider the case where the shock from a supernova is thermalized when it contacts the disk.
We assume that the kinetic energy of the supernova is thermalized throughout the disk.
Here, the energy flux produced in the shock-heated region can be written approximately as follows
\begin{align}
    \mathcal{F}_\mathrm{SN}
    &\sim\left(\cfrac{1}{2}\rho_\mathrm{SN}v_\mathrm{SN}^2\right)v_\mathrm{SN}~ \nonumber\\
    &\approx 3.3\times 10^{7}\,\mathrm{erg\,cm^{-2}\,s^{-1}} 
    \left(\cfrac{E_\mathrm{expl}}{10^{51}\mathrm{~erg}}\right)^{3/2}
    \left(\cfrac{d}{0.1\,\mathrm{pc}}\right)^{-3} ~.
    \label{eq:rad0}
\end{align}
Given that the shock-heated layer (Figure \ref{fig:thermal}) is thin, the radiation from the layer is thought to be released as upward and downward radiation fluxes. 
Additionally, we assume a balance between the downward irradiation flux (heating) and thermal radiation from the disk (cooling). 
The following equations are satisfied:
\begin{align}
    &\mathcal{F}_\mathrm{SN}
    =\mathcal{F}^\uparrow_\mathrm{rad}+\mathcal{F}^\downarrow_\mathrm{rad}
    ~,\label{eq:rad1}  \\
    &\mathcal{F}^\downarrow_\mathrm{rad}\cdot\pi R^2
    = 2\left( \pi R^2\cdot \sigma_\mathrm{SB}T_\mathrm{disk,eq}^4 \right) ~,\label{eq:rad2}
\end{align}
where black-body radiation is assumed as the cooling term for the disk. 
From Eqs.\eqref{eq:rad0}, \eqref{eq:rad1} and \eqref{eq:rad2}, the thermal equilibrium temperature of the disk is given as follows
\begin{align}
    T_\mathrm{disk,eq} 
    =\left( \cfrac{\mathcal{F}^\downarrow_\mathrm{rad}}{8 \sigma_\mathrm{SB}} \right)^{1/4} 
    \approx 735 \,\mathrm{K} 
    \left(\cfrac{E_\mathrm{expl}}{10^{51}\mathrm{~erg}}\right)^{3/8}
    \left(\cfrac{d}{0.1\,\mathrm{pc}}\right)^{-3/4}
    ~.\label{eq:eqtemp}
\end{align}

Finally, we estimate the disk temperature by considering a supernova explosion as a suitable source of \al.
When substituting Eq.\eqref{eq:d-26al} into Eq.\eqref{eq:eqtemp}, we can obtain
\begin{align}
    T_\mathrm{disk,eq} 
    \approx  537 \,\mathrm{K} &
    \left[ 
    \left( \cfrac{\eta_\mathrm{d}}{0.20} \right)
    \left( \cfrac{ M_\mathrm{SN}(\mathrm{^{26}Al}) }{1.3\times 10^{-4}M_\odot} \right)
    \right]^{-3/8} \nonumber \\
    & \times\left( \cfrac{M_\mathrm{disk}}{0.017M_\odot} \right)^{3/8} 
    \left(\cfrac{R}{100\,\mathrm{au}}\right)^{-3/4} 
    ~.\label{eq:main}
\end{align}
The internal energy due to this temperature is well below the gravitational bound energy ($e_\mathrm{th}<e_\mathrm{grav}=GM_\star/R$), so the disk does not thermally evaporate.
However, we find that this disk temperature is an eye-catching value for the evolution of the protosolar disk, and its impact on planet formation cannot be ignored. 
Therefore, we next discuss in Section \ref{subsec:planet} the effect of the temperature we found on planet formation.

\section{Discussion}  \label{sec:discussion}

\add{
\subsection{Effect of inclination angle} \label{subsec:incl}
As in section \ref{sec:model}, this study assumes that the flow from a supernova hits the face-on disk. 
However, in reality, the disk can be
inclined at an angle. }

\add{In terms of injection efficiency, the inclination reduces the contact solid angle, so that Eq \eqref{eq:al-sn} is rewritten as
\begin{align}
    &M_\mathrm{inje}(\mathrm{^{26}Al})
    \sim \eta_\mathrm{d} \cdot \left(\cfrac{\pi R^2\cdot \cos{\theta} }{4\pi d^2}\right) \cdot M_\mathrm{SN}(\mathrm{^{26}Al})_\mathrm{tot} ~,
\end{align}
where $\theta$ is 
the angle between the disk axis and the direction from the solar system to the supernova.
Therefore, the resulting distance required for injection is 
\begin{align}
    d \lesssim 0.20\,\mathrm{pc} 
    &\cdot \cos^{1/2}{\theta} \left[
    \left( \cfrac{\eta_\mathrm{d}}{0.20} \right)
    \left( \cfrac{ M_\mathrm{SN}(\mathrm{^{26}Al})_\mathrm{tot} }{1.3\times 10^{-4}M_\odot} \right)
    \right]^{1/2} \nonumber \\
    &\times\left( \cfrac{M_\mathrm{disk}}{0.017M_\odot} \right)^{-1/2} 
    \left(\cfrac{R}{100\,\mathrm{au}}\right) ~.\label{eq:incl1}
\end{align}
}

\add{
On the other hand, in terms of the disk disruption efficiency, 
the supernova flow that the disk receives per unit area is smaller by $\cos{\theta}$ as well as the injection case.
That is, the stability criterion yields $P_\mathrm{grav} > P_\mathrm{SN}\cdot \cos{\theta}$, and Eq. \eqref{eq:d-press-q} can be rewritten as
\begin{align}
    d >  0.21\,\mathrm{pc}
    &\cdot\cos^{1/3}{\theta}\left(\cfrac{E_\mathrm{expl}}{10^{51}\,\mathrm{erg}}\right)^{1/3}
    \nonumber \\
    &\times
    \left(\cfrac{r}{30\,\mathrm{au}}\right)^{\frac{2+q}{3}}
    \left(\cfrac{M_\mathrm{disk}}{0.017\,M_\odot}\right)^{-1/3}
    ~.  \label{eq:incl2}
\end{align}
}

\add{
The conditions required for injection in Eq. \eqref{eq:incl1} become more strict with $\cos^{1/2}\theta$, whereas the conditions for avoiding disk disruption in Eq. \eqref{eq:incl2} are relaxed with $\cos^{1/3}\theta$.
A simple comparison reveals that a larger inclination angle makes SN disk injection more difficult. 
In other words, 
the assumption of a face-on disk gives the most conservative condition for the injection scenario. 
This supports the overall conclusion that the disk injection scenario is unlikely to work.
}

\subsection{Observational constraints on large dust masses of supernovae} \label{subsec:large-dust}

Dust formation in supernova ejecta, identifiable by infrared (IR) excess, has been confirmed in SN 1987A \citep{2011Sci...333.1258M,2015MNRAS.446.2089W}. 
Considering that SN flows take decades to reach protosolar disks, SN 1987A, about 40 years after the explosion, serves as an excellent case to study dust formation in supernovae relevant to our research.

\cite{2011Sci...333.1258M} observed far-infrared and submillimeter emission from SN 1987A. 
The shapes of the spectral energy distributions (SEDs) were consistent with continuous dust emission, indicating the presence of $0.4-0.7 M_\odot$ of dust. 
This was further supported by extrapolating the synchrotron flux measured at shorter wavelengths up to 8014 days after the explosion to far-infrared wavelengths, which showed an order of magnitude lower emission than observed, indicating dominant dust emission. 
Given the total metal mass of SN 1987A of $\sim3M_\odot$, a dust fraction $\eta_\mathrm{d}\sim20\%$ seems reasonable.

Further research by \citet{2015MNRAS.446.2089W} over 24 years, starting 615 days after the explosion, found $0.6-0.8M_\odot$ of dust, consisting of grains larger than $\gtrsim2 \mathrm{\mu m}$, exists in SN 1987A, suggesting that most of the small dust was formed early and grew through accretion and aggregation. 
These results underscore the importance of considering supernovae at appropriate post-explosion timescales for understanding dust formation rates relevant to solar system formation.

More recently, mid-infrared imaging of SN 1980K over 40 years post-explosion by \citet{2023arXiv231003448Z} reported its dust mass $M_d\approx0.02M_\odot$.
However, the SED analysis indicates that a much greater amount of dust $(\sim 0.24-0.58M_\odot)$, suggesting a dust mass fraction of $\eta_\mathrm{d}\lesssim 0.20$. 

To summarize, for a supernova within the timescale of our study, a dust mass fraction $\eta_\mathrm{d}$ of up to $0.20$ seems realistic. 
This assumes spherical symmetry and non-extreme conditions, but higher $\eta_\mathrm{d}$ values are possible in clumpy structures. 
To understand further details, comprehensive studies are needed, including simulations of multidimensional supernova explosions with dust formation and observations of dust emission in supernovae several decades after the explosion.


\subsection{Effect of disk temperature on planet formation} \label{subsec:planet}

Heating of a protoplanetary disk due to a nearby supernova (Section \ref{sec:thermal}) may cause sublimation and/or melting of materials in the disk and, consequently, influence planet formation. Moreover, such heating may not be consistent with the early solar-system record, as we explain below.

Organic matter on dust grains in the protosolar disk can be irreversibly lost due to the heating. While amorphous carbon component survives at temperatures up to $\sim1000\ \mathrm{K}$, organic materials start to be removed due to pyrolysis and evaporation at $\simeq 250$--$400\ \mathrm{K}$ \citep{2017A&A...606A..16G}. Loss of organic mantles significantly reduces the stickiness of dust grains and thus may limit rocky planet(esimal) formation \citep{2019ApJ...877..128H}. Although the origins of organic matter in carbonaceous chondrites and in returned samples of primitive asteroids in the solar system are poorly understood, some organic matter is considered to trace back to the interstellar medium \citep{glavin2018origin}. Thus, the presence of these pristine materials may put a limit on the heating of the early solar system due to supernova exposure.

Sublimation of highly volatile ices (here defined as materials whose sublimation temperature is lower than water ice) may impose a much lower limit on the temperature experienced during the supernova heating. For instance, major volatile molecules, H$_2$O and CO sublimates at $\simeq 150\ \mathrm{K}$ and $\simeq 20\ \mathrm{K}$, respectively \citep{2016ApJ...821...82O}. Sublimation is a reversible process, and thus, re-condensation after the cooling of the disk may recover its original state. However, sublimation and re-condensation may turn pristine amorphous ice into crystalline ice; this may not be consistent with the property of cometary ice, which is widely thought to be amorphous \citep[e.g.,][]{2020SSRv..216..102R,prialnik2022amorphous}. Moreover, isotopic compositions of highly volatile elements in a comet 67P/Churyumov-Gerasimenko measured with the Rosetta spacecraft suggest that cometary ice has never sublimated and been mixed with the inner solar-system materials \citep{marty2017xenon,2020SSRv..216..102R}. Thus, depending on the re-condensation processes and the efficiency of mixing in the protosolar disk, the heating due to a nearby supernova may be constrained or ruled out.


Lastly, we note that heating to a much higher temperature ($\gtrsim 1300\ {\rm K}$) leads to sublimation and/or melting of silicates. Such high-temperature heating events formed silicate grains, including CAIs and chondrules from their precursor aggregates 
in our solar system \citep[e.g.,][]{2009GeCoA..73.4963K}, which likely changed their stickiness and aerodynamic properties and influenced planet formation; for instance, efficient accretion of chondrules onto large planetesimals is proposed to have assisted forming planetary embryos \citep{2015SciA....1E0109J}.

To summarize, the supernova heating may cause irreversible changes in protoplanetary-disk materials and thus influence planet formation processes. Heating of the protosolar disk associated with implantation of \al\, from a supernova (Section \ref{sec:thermal}) would have caused sublimation of highly volatile ices and, possibly, processing of organic matter, which may contradict the solar system record. Thus, we suggest that future studies need to perform a more detailed analysis of the heating and cooling processes of disks during supernova exposure and physicochemical impacts on the disk materials.

\subsection{Implications for exoplanetary systems} \label{subsec:exoplanets}

Provided that injection into protoplanetary disks is a major source of SLRs in planetary systems, our results indicate that systems like our solar system, whose planet(esimal) formation and evolution are highly influenced by the radioactive decay energy of \al, may be rare. As shown in Section \ref{sec:criteria}, a parameter space that results in injection of \al\, comparable to the amount found in the solar system without disrupting the disk, is fairly limited. In other words, a typical outcome of supernova exposure is either limited injection of \al\,, or disruption of the protoplanetary disk, which implies that existing exoplanetary systems are formed without a significant (solar-system-like) amount of \al. It is not fully understood and beyond the scope of this study how the deficit of \al\, in the protoplanetary disk changes the architecture of resulting planetary systems, but a proposed outcome \citep{2019NatAs...3..307L} is limited loss of water from planetesimals due to insufficient \al\, heat budget and the dominant formation of water-rich planets. Thus, our study ultimately suggests that Earth-like, water-poor planets may be minor in our universe.

\section{Summary} \label{sec:summary}

In this paper, we have assumed that \al\, from a supernova is injected into an already formed protosolar disk and investigated whether there are conditions under which the surviving disk radius can capture enough amount of \al\, for planet formation while allowing for some disk distruption.
We consider a diversity of disk mass and its structure, \al\, yields of SN, and  large-dust mass fraction.
In our model, given the disk mass and its structure, the disk radius that can accept \al\, is obtained without any other assumptions.
We also obtain the position at which the disk is disrupted by ram pressure, depending on the distance from the supernova to the disk.
Under each of these being determined self-consistently, 
we conclude that, as shown in Eqs. \eqref{eq:result} and \eqref{eq:result-al}, an ejecta mass of \al\, is required as follows:
\begin{align}
    M_\mathrm{SN}(\mathrm{^{26}Al})_\mathrm{tot} 
    \gtrsim 
    2.1\times 10^{-3}M_\odot    
    \left( \cfrac{\eta_\mathrm{d}}{0.20} \right)^{-1} 
    \left(\cfrac{E_\mathrm{expl}}{10^{51}\,\mathrm{erg}}\right)^{2/3}
    ~.\nonumber 
\end{align}
This value is difficult to reproduce given the diversity of ejected \al\, mass, and large dust mass fractions from the supernovae, which are shown in Figure \ref{fig:summary}.

Furthermore, we find that even if the above conditions are achieved, the SN shock changes the disk temperature.
And this temperature change is not negligible in the context of planetary formation.

Our finding places a strong constraint on the `disk injection scenario'; a scenario in which fresh-\al\, of the early solar system is injected from a supernova into an already formed protosolar disk is quite challenging.
We rather suggest that the  fresh-\al\, of the early solar system, should have been synthesized/injected in other ways.

\section*{Acknowledgments}
We thank T. Suzuki, S. Inutsuka, K. Maeda, S. Arakawa, and E. Kokubo for fruitful discussion.
This work has been supported by Japan Society for the
Promotion of Science (JSPS) KAKENHI grants 
(18H05437, 20H00174, 20H01904, 20KK0080, 21H04514, 21K13964, 21K13976, 21K13983, 22KJ0528, 22H01290, 22H04571, 22H05150).

\bibliography{ref}{}
\bibliographystyle{aasjournal}

\end{document}